\documentclass[12pt]{article}
\usepackage{times}
\usepackage{geometry}
\usepackage[utf8]{inputenc}
\usepackage{enumitem,amssymb}
\usepackage{ragged2e}
\usepackage{deluxetable}
\usepackage{natbib} 
\usepackage{aas_macros} 
\usepackage{graphicx} 
\usepackage{wrapfig}
\usepackage{framed}

\newlist{thematic}{itemize}{8}
\setlist[thematic]{label=$\square$}
\usepackage{pifont}
\newcommand{\cmark}{\ding{51}}%
\newcommand{\done}{\rlap{$\square$}{\raisebox{2pt}{\large\hspace{1pt}\cmark}}%
\hspace{-2.5pt}}

\begin{document}
\raggedright
\huge
Astro2020 Science White Paper \linebreak

From Stars to Compact Objects: The Initial-Final Mass Relation \linebreak
\normalsize

\noindent \textbf{Thematic Areas:} \hspace*{60pt} $\square$ Planetary Systems \hspace*{10pt} $\square$ Star and Planet Formation \hspace*{20pt}\linebreak
\done Formation and Evolution of Compact Objects \hspace*{34pt} $\square$ Cosmology and Fundamental Physics \linebreak
  \done  Stars and Stellar Evolution \hspace*{8pt} $\square$ Resolved Stellar Populations and their Environments \hspace*{40pt} \linebreak
  $\square$    Galaxy Evolution   \hspace*{49pt} $\square$             Multi-Messenger Astronomy and Astrophysics \hspace*{65pt} \linebreak
  
\textbf{Principal Author:}

Name:	Jessica R. Lu
 \linebreak						
Institution:  UC Berkeley
 \linebreak
Email: jlu.astro@berkeley.edu
 \linebreak
Phone:  310-709-0471
 \linebreak
 
\textbf{Co-authors:} (names and institutions)
  \linebreak
Casey Lam (UC Berkeley, casey\_lam@berkeley.edu), 
Will Dawson (LLNL, dawson29@llnl.gov), 
B. Scott Gaudi, (Ohio State University, gaudi@astronomy.ohio-state.edu),
Nathan Golovich (LLNL, golovich1@llnl.gov),
Michael Medford (UC Berkeley, michaelmedford@berkeley.edu), 
Fatima Abdurrahman (UC Berkeley, fatima.abdurrahman@berkeley.edu),
Rachael L. Beaton (Princeton University, rachael.l.beaton@gmail.com)
  \linebreak

\textbf{Abstract:}
One of the key phases of stellar evolution that remains poorly understood is {\em stellar death}. 
We lack a predictive model for how a star of a given mass explodes and what kind of remnant it leaves behind (i.e. the initial-final mass relation, IFMR). Progress has been limited due to the difficulty in finding and weighing black holes and neutron stars in large numbers. Technological advances that allow for sub-milliarcsecond astrometry in crowded fields have opened a new window for finding black holes and neutron stars: astrometric gravitational lensing. Finding and weighing a sample of compact objects with astrometric microlensing will allow us to place some of the first constraints on the present-day mass function of isolated black holes and neutron stars, their multiplicity, and their kick velocities. All of these are fundamental inputs into understanding the death phase of stellar evolution, improving supernovae models, and interpreting LIGO detections in an astrophysical context. To achieve these goals, we require large area surveys, such as the WFIRST exoplanet microlensing survey, to photometrically identify long-duration ($>$120 day), un-blended microlensing events as candidate compact objects.  We also require high-precision astrometric follow-up monitoring using extremely large telescopes, equipped with adaptive optics, such as TMT and GMT.

\begin{figure}[h]
\begin{framed}
    \begin{minipage}{3.5in}
        \hspace{-0.15in}
        \includegraphics[scale=0.33]{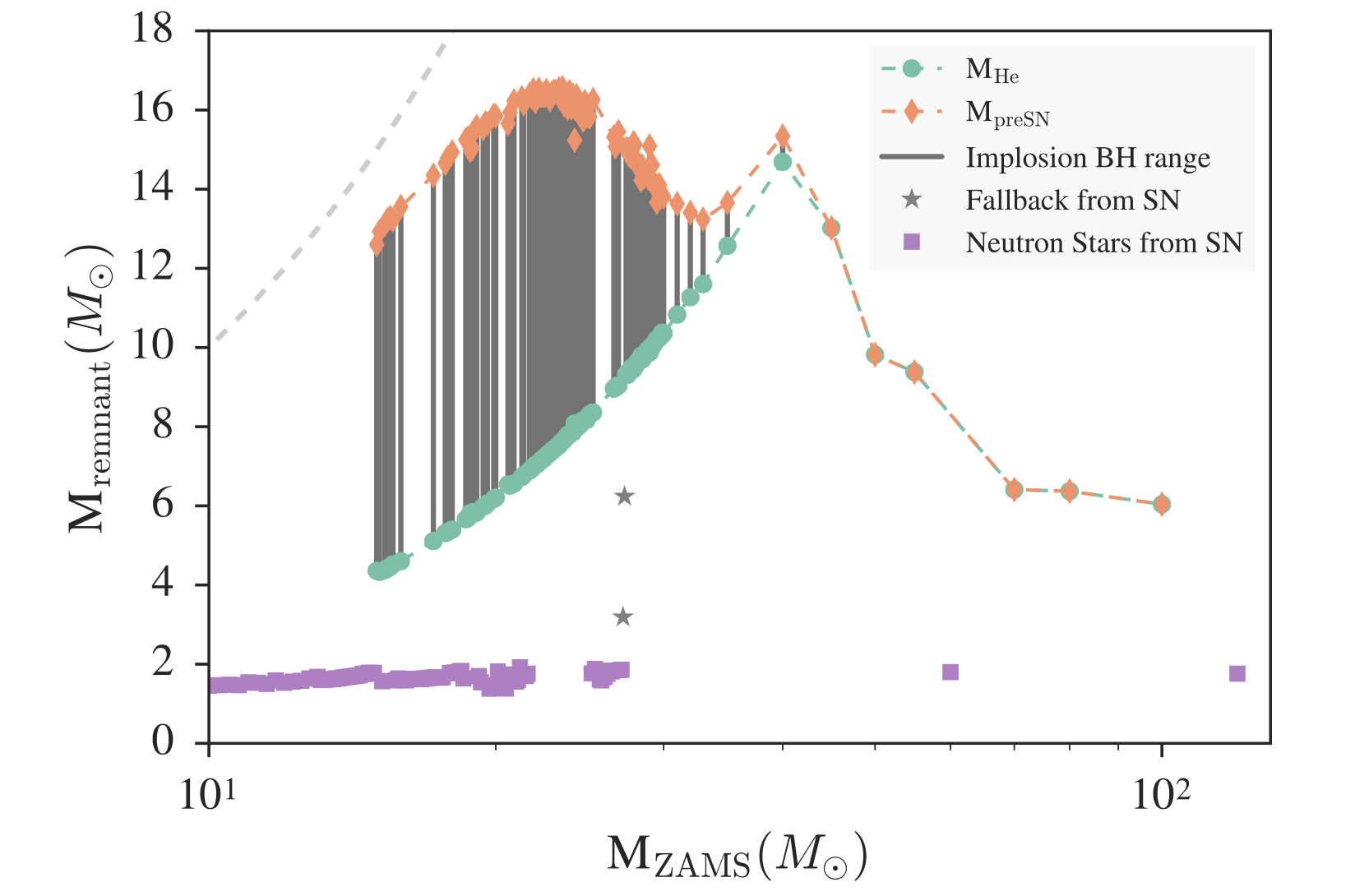}
    \end{minipage}
    \begin{minipage}{2.6in}
    \caption{{\bf Advances in the Theoretical IFMR:} Only in the last 2 years have simulations of supernovae begun to predict a quantitative initial-final mass relation (IFMR). Current predictions produce a probabilistic relationship between the initial stellar mass and the final black hole or neutron star mass. Mass measurements for a sample of black holes and neutron stars are needed to test these models. Figure reproduced from \citep{Raithel:2018}.
    \vspace{0.2in}
    \label{fig:ifmr_theory}
    }
    \end{minipage}
\end{framed}
\end{figure}

\pagebreak
We experience the Universe through the light of stars and they act as a ``fundamental particle" for much of astrophysics. 
While the bulk of stellar evolution is broadly understood, there is a key phase where we have gaping holes in our knowledge: stellar death.
This is one of the most influential periods in a star's life, where it spreads heavy elements and injects energy, sometimes violently through supernovae, into its surroundings. 
The impact of a star's death is closely tied to its mass and we do not have a clear mapping between the initial mass of a star, how it dies, and the type and mass of compact object that results (Figure \ref{fig:ifmr_theory}). Mapping this initial-final mass relation (IFMR) is challenging simply because black holes (BH) and neutron stars (NS) are very difficult to find.  
The type of compact remnant left behind is not only a function of the star's initial ZAMS mass, but also its metallicity \citep{Heger:2003}.
Other more complex parameterizations have also been proposed; see \cite{Ertl:2016} for an example.

\vspace{-0.2in}
Uncertainty in the IFMR has led to order-of-magnitude differences in predictions for the simplest term: the number of BHs in the Galaxy \citep{Agol:2002}. 
Other statistics, such as the BH and NS binary fraction, mass function, and kick velocity distribution, are so uncertain that it is difficult to put current LIGO discoveries in context \citep{Elbert:2018}. 
Ultimately, we need to find and measure masses of a large number of compact objects, which is now possible using gravitational microlensing when a dark lens passes in front of a background star. 

\begin{wrapfigure}[11]{r}{0.35\textwidth}
\vspace{-0.4in}
\begin{framed}
    \includegraphics[scale=0.20]{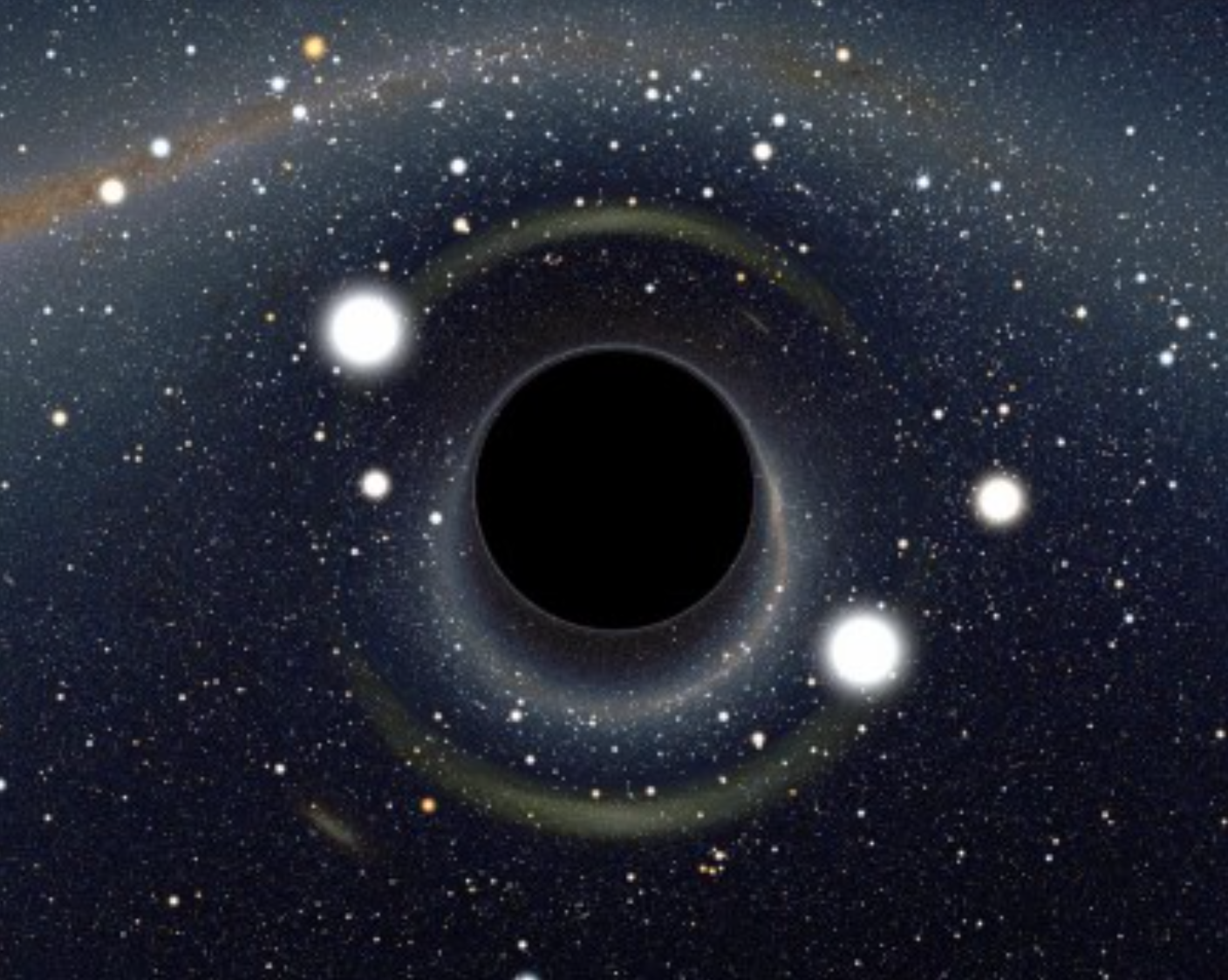}      
    \vspace{-0.2in}
    \caption{
        {\bf Black Holes Lens Background Stars} 
        \label{fig:bh}
    }
\end{framed}
\end{wrapfigure}
\begin{figure}[h]
\begin{framed}
    \centering
    \includegraphics[scale=0.37]{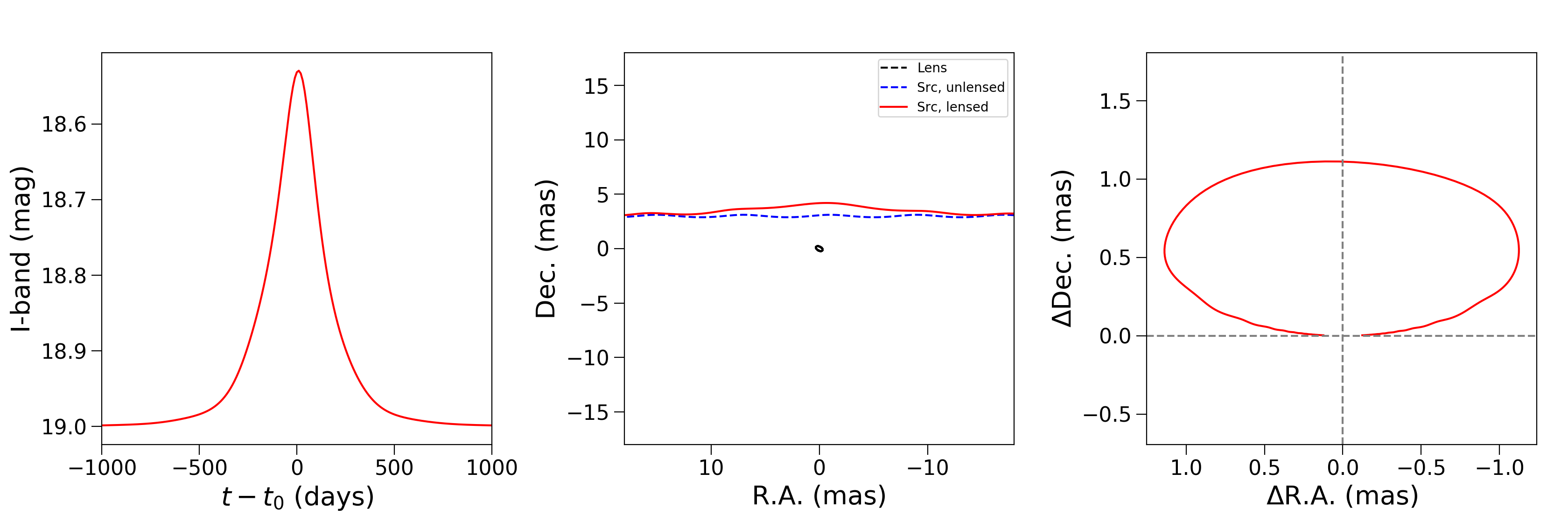}
    \vspace{-0.3in}
    \caption{
    {\bf Photometric and Astrometric Microlensing: } Photometry and astrometry for a black hole at 3 kpc lensing a background star at 6 kpc with a relative proper motion of 8 mas yr$^{-1}$. {\em Left:} Photometric light-curve. {\em Center:} Astrometry of the lens and source, including parallax, as would be seen on the sky. {\em Right:} Astrometry of the lensed source after the proper motion is removed. }
    \label{fig:astrom_lens}
\end{framed}
\end{figure}
\begin{wrapfigure}[14]{r}{0.47\textwidth}
\vspace{-0.5in}
\begin{framed}
    \vspace{-0.1in}
    \includegraphics[scale=0.35]{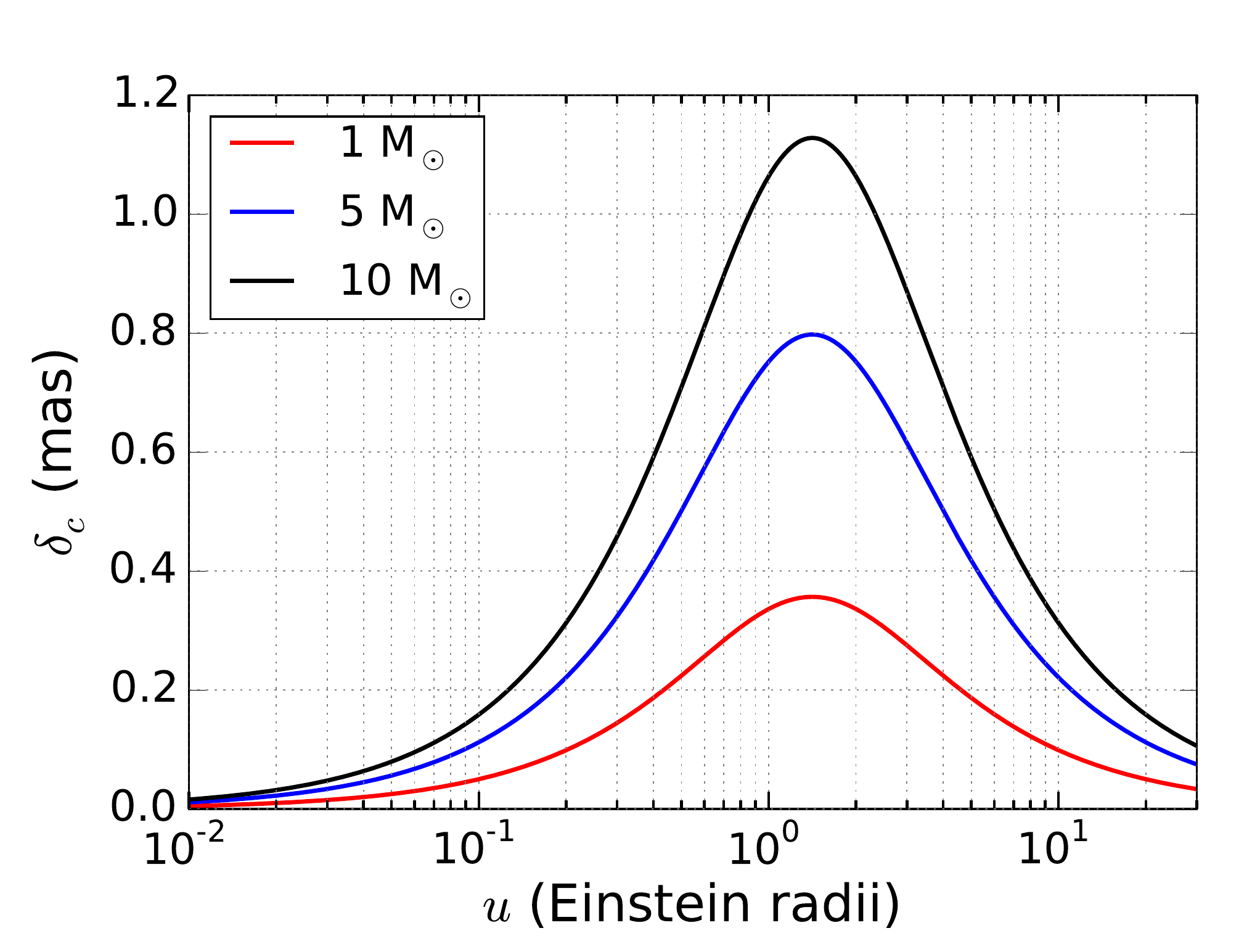}
    \vspace{-0.35in}
    \caption{{\bf Lens Masses from Astrometry:} Astrometric lensing signal vs. separation (or time) for different mass lenses.}
    \label{fig:astrom_mass}
\end{framed}
\end{wrapfigure}

\vspace{0.1in}
A novel approach to detecting low-luminosity objects such as black holes and neutron stars is to use {\bf gravitational microlensing} (Figure \ref{fig:bh}). Black holes or neutron stars that gravitationally lens background stars produce a photometric magnification that has a long duration ($>$1 month, Figure \ref{fig:astrom_lens}). However, a chance alignment of two slow-moving normal stars has a similar photometric signal. 
Fortunately, dark lenses also induce an astrometric shift in the apparent position of the background star that can be $>$0.3 mas \citep{Gould:2014,Lu:2016}. This astrometric shift scales as $M^{1/2}_{lens}$ and, if measured, it allows you to weigh dark lenses (Figure \ref{fig:astrom_mass}).

\vspace{-0.25in}
{\bf Only the combination of photometry and astrometry can be used to estimate the mass of the lens and determine if it is indeed a compact object}. Fortunately, there is an abundance of wide-field photometric surveys planned for the next 10-20 years, including LSST and WFIRST, that will increase the number of dark lens candidates by a factor of 10-100. Current telescopes lack the astrometric precision to measure the masses for this large number of candidates and find the black holes and neutron stars. The next generation extremely large telescopes (ELTs) are ideal, delivering astrometric precisions as good as 15 $\mu$as on faint stars even in crowded regions where Gaia, HST, WFIRST, and JWST, are insufficient. Thus, when a candidate dark lens is identified in a photometric survey, one of the ELTs can be triggered within a few days to observe the astrometric signal and get the lens mass. Also, the $\times$15 mas resolution of the ELTs will allow us to further confirm dark lenses after they move away from the background star in just a few years.

\vspace{0.1in}
{\em Targets:} Microlensing signals are maximized 
when there is a crowded field of background source stars and a high density of foreground objects to act as lenses. The regions that are most suitable for
lensing studies of compact objects created during during 
the death of a star are towards the Milky Way Bulge, which is visible both from the Northern and Southern hemispheres. Large area surveys are required as the microlensing event rate range from $10^{-4} - 10^{-6}$ events star$^{-1}$ yr$^{-1}$. 
Searches for primordial black holes or black holes born from Pop III stars is towards M31 (North) and the LMC/SMC (South). Coverage of both hemispheres is essential. 

\begin{figure}[h]
\begin{framed}
    \centering
    \includegraphics[scale=0.53]{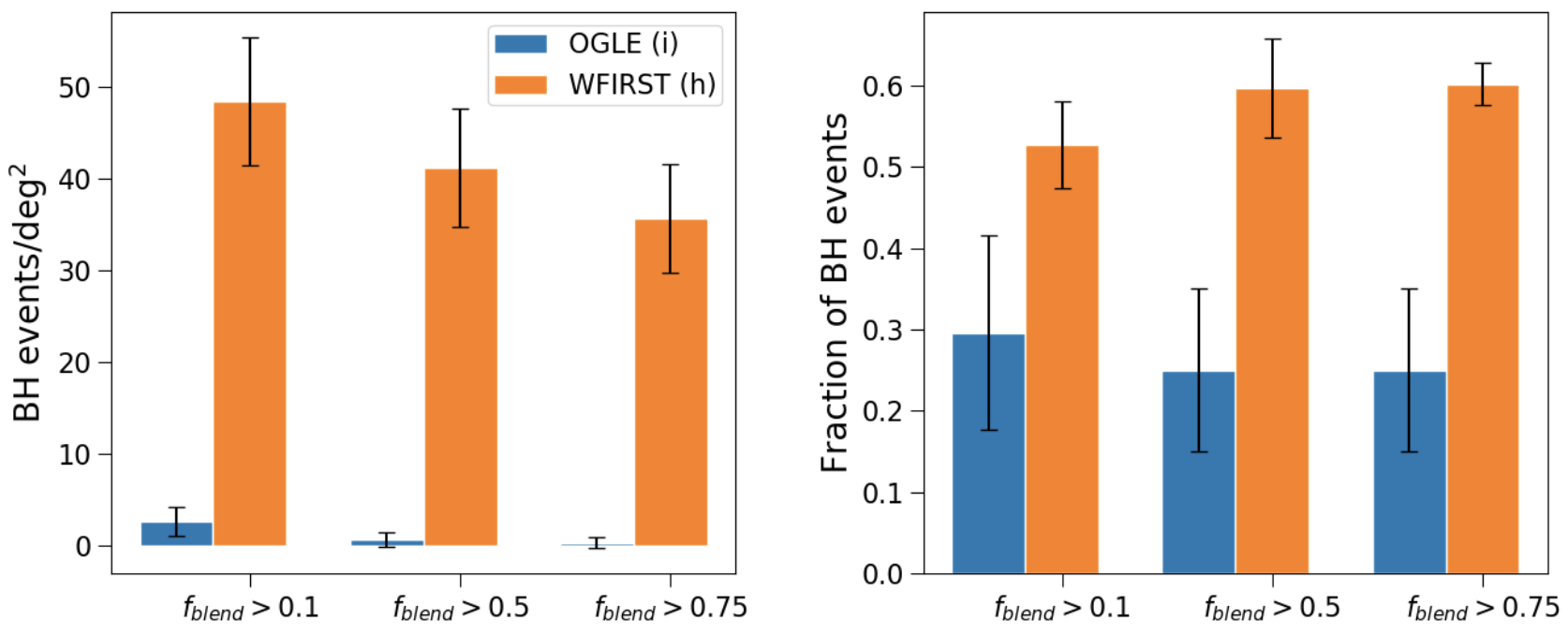}
    \caption{{\bf OGLE vs. WFIRST Identification of Black Hole Candidates: }
    WFIRST significantly increases the number density of black hole lensing events ({\em left}) and our confidence that long-duration, un-blended events are indeed black holes ({\em right}). This is thanks to WFIRST's higher spatial resolution and greater sensitivity at IR wavelengths, where it can see through the obscuring dust towards the crowded Milky Way Bulge. These simulations were produced with the Population Synthesis of Compact Object Lensing Events (PopSyCLE, Lam in prep.).
    \label{fig:wfirst_ogle}
    }
\end{framed}
\end{figure}

\vspace{0.1in}
{\bf Required Facilities:} Microlensing events are first detected photometrically in wide-field photometric surveys such as OGLE, MOA, LSST, and WFIRST. Dedicated microlensing surveys, such as the WFIRST exoplanet survey, will also find numerous compact objects so long as the duration of the survey is long. Microlensing events by black holes have timescales that can extend up to 600 days; thus a survey length of $>2$ yr is needed. WFIRST is also the only facility that has high spatial resolution during the detection phase, which reduces confusion and dramatically increases the probability of correctly identifying black hole candidates in advance (Figure \ref{fig:wfirst_ogle}).

Once good compact object candidates are identified, they must be monitored astrometrically. The astrometric follow-up is demanding as it requires high sensitivity, high-precision astrometry, and high spatial resolution, which is just at the boundary of what the Keck adpative optics system can do today and is not possible with Gaia (too shallow in crowded regions). While WFIRST will provide astrometry at the $0.5$ mas level, the future ELTs, equipped with adaptive optics (AO), will have an astrometric precision at least an order of magnitude better and spatial resolutions of $<$20 mas.  Furthermore, simultaneous measurements with a ground-based ELT and WFIRST in an L2 orbit will enable instantaneous space parallax measurements, which constrains the distance ratio between the lens and source. 

\vspace{0.1in}
ELT follow-up observations of compact object candidates must be flexible and prompt as events are only clearly identified as black hole candidates within 1-2 weeks of the photometric peak. High-precision astrometric observations are required in a single filter over multiple years. Typically, each target must be observed 4-5 times in the first year and 2-3 times for the following 3 years, totalling 10-15 measurements for each target.  Additional single-epoch observations in other filters are helpful for testing binary lens or source scenarios.  
The Thirty Meter Telescope (TMT) has the advantage with a larger aperture and also the gravity-invariant IRIS imager. However, many BHs and neutron stars may be in binary systems, which can affect the mass measurement, and high-resolution spectroscopy (R$\sim$50,000) at the diffraction limit of the ELTs, as is available on the Giant Magellan Telescope's GMT-NIRS instrument, can easily detect the binary from orbital motion. 
For astrometry, in order to reach the desired 50 $\mu$as precision, a SNR$>$300 is needed in the AO images. Typically, the astrometric precision is set by systematic errors such as lack of knowledge of the PSF and variable distortion. So it is important that the future ELTs {\em deliver astrometric calibrations and PSF reconstruction.} 

\vspace{0.1in}
Ultimately, the ELT data is combined with well-sampled photometric light curves from WFIRST or LSST. The astrometry and photometry from all facilities is jointly fit to microlensing models. Development of detailed microlensing modeling codes as well as population synthesis will be essential to interpret the observations and constrain the IFMR.

\vspace{0.1in}
{\bf Summary of Recommendations}
\begin{itemize}
    \item WFIRST microlensing survey $>$2 yr
    \item ELT (i.e. TMT and GMT) equipped with adaptive optics
    \item Astrometric distortion calibrations for WFIRST and ELTs
    \item PSF modeling and reconstruction for WFIRST and ELTs
    \item Funding support for long-duration astrometric experiments as well as theoretical work in microlens modeling and population synthesis of compact objects. 
\end{itemize}

\pagebreak

\vspace{0.1in}
\bibliography{imf_ifmr}

\end{document}